\documentclass[12pt]{article}
\usepackage[english]{babel}
\usepackage{amsmath,amssymb,algorithm,algorithmic,bbm,multirow,mathtools,booktabs,amsthm} 

\usepackage[utf8]{inputenc}
\usepackage{comment}

\usepackage{float}

\usepackage{amsfonts}

\usepackage{tabularx}
\usepackage{multirow}
\usepackage{tabu}
\usepackage{makecell}
\usepackage{threeparttable}
\usepackage{epsfig, graphicx, graphics}
\usepackage[dvipsnames]{xcolor}

\usepackage{mathtools}
\usepackage{hhline}
\usepackage{acro}
\usepackage[switch]{lineno}

\usepackage[long]{optidef}
\usepackage{booktabs}
\usepackage{adjustbox}

\usepackage{natbib}
\usepackage[colorlinks = true,
            linkcolor = blue,
            urlcolor  = blue,
            citecolor = blue,
            anchorcolor = blue]{hyperref}
\usepackage{etoolbox}

\newtheorem{theorem}{Theorem}[section]

\begin{document}

\def\spacingset#1{\renewcommand{\baselinestretch}%
{#1}\small\normalsize} \spacingset{1}


\title{\bf Robust Sparse Bayesian Infinite Factor Models}
\author{Jaejoon Lee \\
Department of Statistics, Seoul National University\\
and \\
Jaeyong Lee \\
Department of Statistics, Seoul National University}
\maketitle

\bigskip
\begin{abstract}
Most of previous works and applications of Bayesian factor model have assumed the normal likelihood regardless of its validity. We propose a Bayesian factor model for heavy-tailed high-dimensional data  based on multivariate Student-$t$ likelihood to obtain better covariance estimation. We use multiplicative gamma process shrinkage prior and factor number adaptation scheme proposed in Bhattacharya \& Dunson [\textit{Biometrika} (\citeyear{bhattacharya2011sparse}) 291--306]. Since a naive Gibbs sampler for the proposed model suffers from slow mixing, we propose a Markov Chain Monte Carlo algorithm where fast mixing of Hamiltonian Monte Carlo is exploited for some parameters in proposed model. Simulation results illustrate the gain in performance of covariance estimation for heavy-tailed high-dimensional data. We also provide a theoretical result that the posterior of the proposed model is weakly consistent under reasonable conditions. We conclude the paper with the application of proposed factor model on breast cancer metastasis prediction given DNA signature data of cancer cell.
\end{abstract}
\noindent%

\newpage
\spacingset{1.5} 
\section{Introduction}
\label{intro}
Factor model is a highly efficient tool to understand the covariance structure of high-dimensional  data. The covariance structure is captured by representing the $p$-dimensional observation as the sum of linear transformation of latent factors ($k \ll p$) and an error term. In the factor model, the covariance matrix $\Omega$ has the form of $\Omega = \Lambda\Lambda^T + \Sigma$, where $\Lambda$ is a $p\times k$ factor loading and $\Sigma$ is a $p\times p$ diagonal error variance matrix. Due to the parsimony of representing $p\times p$ covariance with only $p(k+1)$-dimensional parameters, the factor model is widely used for covariance estimation in many applications with high-dimensional data, e.g. spatial analysis \citep{lopes2008spatial} and genomics \citep{carvalho2008high}.

The number of latent factors $k$ is a key element in the factor model. Variations of the factor model have been proposed for the estimation of the number of factors. \citet{lopes2004bayesian} updated the number of latent factors in the posterior sampling process using reversible jump Markov Chain Monte Carlo \citep{green1995reversible}. \citet{ando2009bayesian} determined the number of latent factors by maximizing the marginal likelihood, which is analytically derived with a chosen prior distribution. \citet{bhattacharya2011sparse} proposed \textit{multiplicative gamma process shrinkage prior}, which is a prior for the infinite factor model and encourages factor loadings with large indices to be close to 0. In the posterior sampling, the number of factors is adapted by adding or deleting latent factors depending on the sparsity of the current factor loading estimate. Such adaptation in \citet{bhattacharya2011sparse} is desirable in that an additional calculation is not required. Moreover, it is guaranteed that the Markov Chain Monte Carlo (MCMC) algorithm using factor adaptation is ergodic.  

For the last few decades, many approaches have been made to obtain sparse estimator under the high-dimensional setting. Variations of factor model have been proposed in a similar vein. \citet{west2003bayesian} and \citet{carvalho2008high} used the spike-and-slab prior on factor loadings, which is a mixture of a point mass at $0$ and a continuous density. Although the point mass mixture prior is intuitive and does induce sparse estimates, it has a critical disadvantage of slow mixing and convergence. Later, due to the advantage in posterior computation over point mass mixture prior, factor models using global-local shrinkage prior \citep{polson2010shrink} have been suggested. For example, the aforementioned infinite factor model of \citet{bhattacharya2011sparse} assigned multiplicative gamma process shrinkage prior on factor loadings, and \citet{ferrari2020bayesian} proposed a factor regression model using Dirichlet-Laplace shrinkage prior \citep{bhattacharya2015dirichlet} on factor loadings.

Most of the factor models aforementioned are based on the normality assumption which, however, is ill-suited when outliers are present. \citet{ando2009bayesian} proposed a factor model with matrix-variate $t$ distribution to obtain robust estimate. \citet{zhang2014robust} proposed a robust version of the factor model utilizing the fact that a multivariate $t$ distribution can be represented as a scale mixture of normal distributions. To the best of our knowledge, however, no approach has been proposed for both robustness against outliers and sparsity of the estimate. 

This work proposes a \textit{robust sparse Bayesian infinite factor model}, which estimates covariance robustly under heavy tail distribution. Specifically, it is an extension of the \textit{sparse Bayesian infinite factor model} \citep{bhattacharya2011sparse}, utilizing the multivariate $t$ likelihood instead of normal likelihood. Under the heavy tail distribution, the proposed model has improved performance of covariance estimation over the normal-likelihood factor model of \citet{bhattacharya2011sparse}. Also, we show that under the assumption of known degrees of freedom of $t$-distribution, the posterior is consistent under the weak topology. Despite the optimal value of $t$ degrees of freedom is not given in the real data analysis, simulation results indicate that the proposed model outperforms the normal-likelihood factor model by choosing a sufficiently small number as the degrees of freedom of $t$ distribution.

In Sect.~\ref{sec:2} the sparse Bayesian infinite factor model of \citet{bhattacharya2011sparse} is introduced. In Sect.~\ref{sec:3} we propose robust factor model with Student's $t$-likelihood. The posterior computation algorithm and theoretical properties are also presented. In Sect.~\ref{sec:4} performance of the proposed model is demonstrated through simulation studies. In Sect.~\ref{sec:5} the proposed model is applied to prediction of breast carcinoma metastasis using microarray data of cancer tissue. The discussion is given in Sect.~\ref{sec:6}. 

\section{Sparse Bayesian Infinite Factor Models}
\label{sec:2}
The \textit{sparse Bayesian infinite factor model} \citep{bhattacharya2011sparse} is a Bayesian factor model specialized for high-dimensional covariance estimation. The joint distribution of observation $\mathbf y_i \in \mathbb R^p$ and latent factor $\eta_i \in \mathbb R^k$ is as follows:
\begin{equation*}
\begin{bmatrix}
\mathbf y_i \\
\eta_i
\end{bmatrix} \Big\vert \Lambda, \Sigma
\stackrel{iid}{\sim} \mathcal N_{p+k}\left( 
\begin{bmatrix}
\mathbf 0 \\
\mathbf 0
\end{bmatrix},
\begin{bmatrix}
\Lambda \Lambda^T + \Sigma & \Lambda\\
\Lambda^T & \mathbf I_d
\end{bmatrix}
\right)
, \enspace i = 1,2, \ldots, n.
\end{equation*}The model is differentiated from the preceding Bayesian factor models in mainly two points: its expanded parameterization on factor loading $\Lambda$ and the adaptation on the number of factors $k$.

Choosing the number of the latent factor $k$ is an important issue. The model addresses this issue by first allowing the parameter space $\Theta_\Lambda$ to contain all possible numbers of latent factor and by dynamically truncating the insignificant latent factors in posterior computation. The parameter space of factor loading $\Lambda$ and error covariance $\Sigma$ are as follows:
\begin{equation*}
\Theta_\Lambda = \left\{ \Lambda = (\lambda_{jh}), j= 1, \ldots, p, h=1, \ldots, \infty, \max_{1\leq j\leq p}\sum_{h=1}^\infty \lambda_{jh}^2 < \infty \right\},
\end{equation*}
\begin{equation*}
\Theta_\Sigma = \Bigg\{ \Sigma \in \mathbb R^{p \times p} : \Sigma_{jj} > 0 \enspace \forall j= 1, \ldots, p , \enspace \Sigma_{ij} = 0 \enspace \forall 1\leq i \neq j \leq p  \Bigg\}, 
\end{equation*}
where $\Sigma_{ij}$ is the $(i, j)$th element of matrix $\Sigma$. Note that the condition 
\begin{equation*}
    \max_{1\leq j\leq p}\sum_{h=1}^\infty \lambda_{jh}^2 < \infty
\end{equation*}
is a necessary and sufficient condition for all the entries of $\Lambda \Lambda^T$ to be finite so that the resulting covariance matrix $\Omega = \Lambda \Lambda^T + \Sigma$ is defined.

For prior $\Pi_\Lambda$ on factor loadings with infinitely many latent factors, \textit{the multiplicative gamma process prior} is proposed. It is a global-local shrinkage prior \citep{polson2010shrink} having entry-wise and column-wise variance components as local and global variance components, respectively. Also, choosing $a_2 \geq 1$, it is designed so that the strong shrinkage is imposed for the factors with large column index. The full prior specification of sparse Bayesian infinite factor models is as follows:
\begin{equation}
\begin{split}
&\lambda_{jh} \vert \phi_{jh}, \tau_h \sim \mathcal N( 0, \phi_{jh}^{-1} \tau_h^{-1}),  \enspace \phi_{jh} \sim \text{Ga}(\kappa/2, \kappa/2),\enspace \tau_{h} = \prod_{l = 1}^h \delta_l, \\
\delta_1 &\sim \text{Ga}(a_1, 1), \enspace \delta_l \sim \text{Ga}(a_2, 1), \enspace l \geq 2, \enspace a_1 \sim \text{Ga}(2,1), \enspace a_2 \sim \text{Ga}(2,1) \label{mgpprior} \\
&\Sigma = \text{diag}(\sigma_1^{-2}, \ldots, \sigma_p^{-2}), \enspace \sigma_j^{-2} \sim \text{Ga}(a_\sigma, b_\sigma), \enspace j = 1, \ldots, p. 
\end{split}
\end{equation}

The number of factors $k$ is determined adaptively by adding or removing latent factor within MCMC iterations, inspecting current factor loading estimate $\hat \Lambda^{(t)}$. At the $t$th iteration of MCMC, the chain goes through adaptation step with decreasing probability $p(t)$, say $p(t) = 1/\exp(1+0.0005t)$. In adaptation step, if there are columns of the current value $\Lambda^{(t)}$ whose entries are all close to zero under prespecified threshold, the columns are removed, otherwise new columns are generated from the prior distribution and are added to the current factor loadings. Also corresponding columns of latent factor matrix $\eta$, variance components $\phi_{jh}, \delta_h$ for deleted (added) column of factor loadings are also deleted (added) accordingly. The adaptation procedure is to keep only the effective latent factors whose factor loadings take up a large part of current posterior sample of the covariance.

This adaptive method has a significant advantage of computation, compared to other methods which needed additional MCMC step \citep{lopes2004bayesian} or comparison of other model selection criteria \citep{ando2009bayesian}. As justification for their adaptation scheme, \citet{bhattacharya2011sparse} showed that, with the prior specified as Eq.~\ref{mgpprior}, the prior probability of approximated covariance $\Omega_H = \Lambda_H^T \Lambda_H + \Sigma$ being arbitrarily close to $\Omega = \Lambda \Lambda^T + \Sigma$ converges to $1$ at exponential rate as $H$ goes to $\infty$, where $\Lambda_H$ is a truncated factor loading of $\Lambda$ with first $H$ columns. Furthermore, the adaptation procedure satisfies the diminishing adaptation condition in \citet{roberts2007coupling}. Thus the convergence of the MCMC algorithm is guaranteed.

\section{Robust Sparse Bayesian Infinite Factor Models}
\label{sec:3}
\subsection{Model}
\label{sec:3.1}
The sparse Bayesian infinite factor model is a factor model based on the normal likelihood. Even though the model has proven its success in high-dimensional covariance estimation, the model may not be the best option when there are outliers in the data or the error distribution has heavy tail. We extend the model by replacing the normal distribution with $t$-distribution which has heavier tail and propose \textit{robust sparse Bayesian infinite factor model}.

\begin{figure}
\centering
\includegraphics[width=0.4\textwidth]{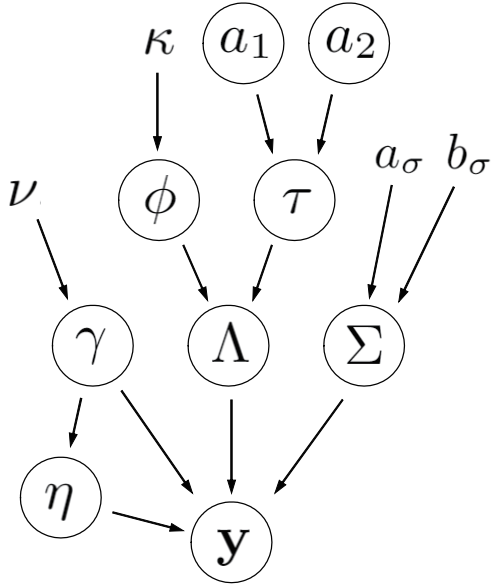}
\caption{The directed acyclic graph representation for the proposed models}
\label{fig:1}
\end{figure}

A multivariate $t$ distribution has a polynomial tail instead of exponential one. The probability density function of multivariate $t$ distribution is as follows:
\begin{equation*}
f(\mathbf y \vert \nu, \mu, \Omega) = \frac{\Gamma( \frac{\nu + p}{2} )}{\Gamma(\frac{\nu}{2})(\nu\pi)^{p/2} \det(\Omega)^{1/2}} \left[ 1 + \frac{(\mathbf y-\mu)^T \Omega^{-1}(\mathbf y-\mu)}{\nu}\right]^{-\frac{\nu + p}{2}},
\end{equation*}
where $\Gamma(x)$ is a gamma function and $\det(A)$ is the determinant of a square matrix $A$. When extending normal likelihood to the $t$ likelihood, we use an equivalent representation of multivariate $t$ distribution as a scale mixture of normal distributions.
\begin{equation*}
\begin{bmatrix}
\mathbf y_i \\
\eta_i
\end{bmatrix} \Big\vert \Lambda, \Sigma, \nu
\stackrel{ind}{\sim} t_{p+k}\left( \nu,
\begin{bmatrix}
\mathbf 0 \\
\mathbf 0
\end{bmatrix},
\begin{bmatrix}
\Lambda \Lambda^T + \Sigma & \Lambda\\
\Lambda^T & \mathbf I_d
\end{bmatrix}
\right)
, \enspace i = 1,2, \ldots, n.
\end{equation*}
\begin{equation}
\begin{split}
\iff
\begin{bmatrix}
\mathbf y_i \\
\eta_i
\end{bmatrix}
\Big\vert \gamma_i, \Lambda, \Sigma
&\stackrel{ind}{\sim} \mathcal N_{p+k}\left( 
\begin{bmatrix}
\mathbf 0 \\
\mathbf 0
\end{bmatrix},
\frac{1}{\gamma_i}
\begin{bmatrix}
\Lambda \Lambda^T + \Sigma & \Lambda\\
\Lambda^T & \mathbf I_d
\end{bmatrix}
\right)
, \enspace i = 1,2, \ldots, n.  \label{normalrepresentation} \\
\gamma_i \vert \nu &\stackrel{iid}{\sim} \text{Ga}\left(\frac{\nu}{2}, \frac{\nu}{2} \right), \enspace i = 1,2,\ldots, n. 
\end{split}
\end{equation}
It is desirable to use the representation in Eq.~\ref{normalrepresentation} because posterior computation is a straightforward Gibbs update, exploiting conjugacy of the normal model with the normal prior distribution. The directed acyclic graph representation for the proposed model is illustrated in Fig.~\ref{fig:1}. The details of the posterior computation of the proposed model are explained in Sect.~\ref{sec:3.2}.

For the prior distribution of factor loading $\Lambda$ and error variance $\Sigma$, we follow the prior $\Pi_\Lambda$ and $\Pi_\Sigma$ as defined in Eq.~\ref{mgpprior}. As we are dealing with multivariate $t$-distribution, we have $\nu$, the degrees of freedom, as an additional parameter. We fix $\nu$ at sufficiently small value in all analyses. From extensive simulation studies, we found that datasets with moderate size have only dim information for $\nu$ and unspecified $\nu$ render slow mixing in the posterior sampling. Thus, the improved performance can be attained by choosing sufficiently small value of $\nu$ in the presence of outliers. The simulation results under different choices of $\nu$ are demonstrated in Sect.~\ref{sec:4}.

\subsection{Inference}
\label{sec:3.2}
While most of the posterior computation steps of the proposed model are similar to those in \citet{bhattacharya2011sparse}, a Gibbs update step can be modified to incorporate the auxiliary variable $\gamma_i$ which extends the normal likelihood to the multivariate Student-$t$ likelihood. For the number of latent factors $k$, we use the same factor adaptation strategy to adaptively determine $k$ as illustrated in Sect.~\ref{sec:2}.

Since all conditionals are tractable distributions, a straightforward Gibbs sampler can be implemented for posterior computation of the proposed factor model with $t$ likelilhood, as in the normal-likelihood factor model of \citet{bhattacharya2011sparse}. However, in high-dimensional setting ($n \ll p$), we have observed slow mixing of Markov chain when naive Gibbs sampler is implemented on the proposed model. To cope with the computational issue arising from more complicated model structure, we made two additional modifications upon Gibbs sampler; collapsing and Hamiltonian Monte Carlo.

The collapsed Gibbs sampler \citep{liu1994collapsed} is a variation of Gibbs sampler which utilizes the conditional of \textit{collapsed} version of joint distribution with some parameters are marginalized out of the condition term. Decoupling some dependencies between conditionals, it is known that the collapsed Gibbs sampler leads to faster mixing than that of the Gibbs sampler. We apply this collapsing idea on $\eta$ and $\gamma$. This is equivalent to regarding $\eta$ and $\gamma$ as a block of single parameter and updating them at a single step of a Gibbs sampler.
\begin{equation*}
p(\gamma_i , \eta_i \vert \mathbf y_i, \cdots)  = p(\eta_i \vert \mathbf y_i, \cdots)p(\gamma_i \vert \eta_i, \mathbf y_i, \cdots)
\end{equation*}
\begin{equation*}
p(\eta_i \vert \mathbf y_i, \cdots) \sim t_k \left( \eta_i : \nu + p, (I+\Lambda \Sigma \Lambda)^{-1}\Lambda \Sigma \mathbf y_i, \frac{\nu + \mathbf y_i^T\mathbf y_i}{\nu + p}(I+\Lambda \Sigma \Lambda)^{-1} \right)
\end{equation*}
\begin{equation*}
p(\gamma_i \vert \eta_i, \mathbf y_i, \cdots) \sim \text{Ga}\left( \gamma_i:\frac{\nu + p + k}{2}, \frac{\nu + (\mathbf y_i- \Lambda \eta_i)^T\Sigma^{-1}(\mathbf y_i- \Lambda \eta_i) + \eta_i^T \eta_i}{2} \right)
\end{equation*}
Fundamentally, the Gibbs sampler is a random-walk Metropolis algorithm with full conditional as a proposal distribution. Both methods explore parameter space via random walk which is highly inefficient for high-dimensional parameter space. Nowadays, in such a case with high-dimensional parameters, the Hamiltonian Monte Carlo is considered to be a gold-standard for posterior computation and has proven empirical success in many applications. The Hamiltonian Monte Carlo uses an auxiliary variable (momentum) and the information from gradient of the log-posterior to perform better search.

To deal with the complicated model structure of $\gamma$ which affects both latent variable $\eta$ and observation $\mathbf y$, we apply No-U-Turn sampler \citep{hoffman2014no} for updating $\eta$. The No-U-Turn sampler is a variation of the Hamiltonian Monte Carlo which automatically tunes the path length of Hamiltonian approximation. Though the No-U-Turn sampler is often used to update all parameters in the model, we applied single No-U-Turn sampler update per iteration. This is comparable to commonly used Metropolis-within-Gibbs scheme, which updates some parameter with Metropolis update while updating the others with Gibbs sampler. Applying No-U-Turn sampler on $n\times k$ dimensional $\eta$, we aim to keep both simplicity of overall posterior computation and better mixing of Hamiltonian Monte Carlo in posterior inference. 

For the Metropolis-Hastings updates of $a_1$ and $a_2$, we used Gaussian proposal with lower bound constraint of $a_1 > 2$ and $a_2 > 3$, respectively. It is a sufficient condition that induced prior on each entry of covariance $\Omega$ has finite second moment. Refer to Sect. 2.2 of \citet{bhattacharya2011sparse} for the detailed explanation. Also \cite{durante2017note} suggests that choosing $a_2$ moderately higher than $a_1$ facilitates better shrinkage of factor loadings, which motivates higher lower bound for $a_2$ than $a_1$. The MCMC algorithm for robust sparse Bayesian infinite factor models given the number of factors $k$ is as follows:
\begin{enumerate}
    \item Sample $\lambda_j$, the $j$th row of factor loading $\Lambda$, for $j=1,\ldots,p$ from normal distribution:
\begin{equation*}
\begin{split}
p(\lambda_j \vert \cdots) \sim \mathcal N_{k}\left( \lambda_j : \Psi_\Lambda^{j} \left(\sigma_j^{-2} \sum_{i=1}^{n} \gamma_i y_{ij} \eta_i\right), \Psi_\Lambda^{j} \right), \\
\text{where }\Psi_\Lambda^{j} = \left(\sigma_j^{-2} \sum_{i=1}^{n} \gamma_i \eta_i \eta_i^T+\text{diag}(\phi_{jh}\tau_h)\right)^{-1}.
\end{split}
\end{equation*}
\item Sample $\sigma_j^{-2}$, for $j=1,\ldots,p$ from gamma distributions:
\begin{equation*}
p(\sigma_j^{-2} \vert \cdots) \sim \text{Ga}\left( \sigma_j^{-2} : a_\sigma + \frac{n}{2}, b_\sigma + \frac{\sum_{i=1}^{n} \gamma_i (y_{ij} - \lambda_j^T \eta_i)^2}{2} \right).
\end{equation*}
\item Sample $\eta_i$, for $i=1,\ldots,n$ with a single iteration of No-U-Turns-Sampler with step size $\epsilon$ from $t$ distribution:
\begin{equation*}
p(\eta_i \vert \mathbf y_i, \cdots ) \sim t_k \left( \eta_i : \nu + p, (I+\Lambda \Sigma \Lambda)^{-1}\Lambda \Sigma \mathbf y_i, \frac{\nu + \mathbf y_i^T\mathbf y_i}{\nu + p}(I+\Lambda \Sigma \Lambda)^{-1} \right).
\end{equation*}
\item Sample $\gamma_i$, for $i=1,\ldots,n$ from gamma distributions:
\begin{equation*}
p(\gamma_i \vert \eta_i, \mathbf y_i, \cdots) \sim \text{Ga}\left( \gamma_i:\frac{\nu + p + k}{2}, \frac{\nu + (\mathbf y_i- \Lambda \eta_i)^T\Sigma^{-1}(\mathbf y_i- \Lambda \eta_i) + \eta_i^T \eta_i}{2} \right).
\end{equation*}
\item Sample $\phi_{jh}$, for $j=1,\ldots,p$, $h = 1, \cdots, k$ from gamma distributions:
\begin{equation*}
p(\phi_{jh} \vert \cdots) \sim \text{Ga}\left(\phi_{jh} : \frac{\kappa + 1}{2}, \frac{\kappa + \tau_h \lambda_{jh}^2 }{2}\right).
\end{equation*}
\item Sample $\delta_h$, for $h=1,\ldots,k$ from gamma distributions:
\begin{equation*}
p(\delta_1 \vert \cdots) \sim \text{Ga}\left( \delta_1 : a_1 + \frac{pk}{2}, 1+\frac{\sum_{\ell=1}^{k} \sum_{j=1}^{p} \tau_\ell  \phi_{j\ell} \lambda_{j\ell}^2}{2} \right),
\end{equation*}
\begin{equation*}
p(\delta_h \vert \cdots) \sim \text{Ga}\left( \delta_h : a_2 + \frac{p(k-h + 1)}{2}, 1+\frac{\sum_{\ell=h}^{k} \sum_{j=1}^{p} \tau_\ell  \phi_{j\ell} \lambda_{j\ell}^2}{2} \right), \enspace h \geq 2.
\end{equation*}
\item Sample $a_1, a_2$ by Metropolis-Hastings update, using Gaussian proposal with lower bound constraint of $a_1 > 2$ and $a_2 > 3$.
\end{enumerate}

\subsection{Theoretical Properties}
\label{sec:3.3}
\citet{bhattacharya2011sparse} showed the weak consistency of the posterior density of their model. In this section, we show that the posterior density of the proposed model is weakly consistent, given that the degrees of freedom $\nu$ of the t-distribution is well-specified. All proofs for theorems can be found in Appendix.

For the sake of coherence, we follow the notation of \citet{bhattacharya2011sparse}. $\Pi_\Lambda$ and $\Pi_\Sigma$ are prior distribution on $\Theta_\Lambda$ and $\Theta_\Sigma$, respectively. $\Theta_\Omega$ is a space of $p\times p$ positive semi-definite matrices, and an open ray $\Theta_\nu = (2,\infty)$ is a parameter space for the degrees of freedom $\nu$. Let $g : \Theta_\Lambda \times \Theta_\Sigma \rightarrow \Theta_\Omega$ be a mapping which maps $(\Lambda, \Sigma)$ to covariance matrix as follows:
\begin{equation*}
g(\Lambda, \Sigma) = \Lambda \Lambda^T + \Sigma.
\end{equation*}
Let $\tilde g: \Theta_\nu \times \Theta_\Lambda \times \Theta_\Sigma \rightarrow \Theta_\nu \times \Theta_\Omega$ be a mapping such that:
\begin{equation*}
\tilde g ((\nu, \Lambda, \Sigma)) = (\nu, g(\Lambda, \Sigma)) = (\nu, \Lambda \Lambda^T + \Sigma).
\end{equation*}
The parameters of multivariate $t$ likelihood are $(\nu, \Omega)$. Then full prior distribuion $\Pi$ on $\Theta_\nu \times \Theta_\Omega$ is $\Pi = (\Pi_\nu \otimes \Pi_\Lambda \otimes \Pi_\Sigma) \circ \tilde g^{-1}$ which is induced by $\Pi_\nu$, $\Pi_\Lambda$, $\Pi_\Sigma$. If we prespecify the degrees of freedom $\nu$, say $\nu = \tilde \nu$, then it is equivalent to choosing $\Pi_\nu$ as a Dirac probability measure at some point $\tilde \nu$.
\begin{theorem}
\label{thm:1}
\textit{
Let
\[
B^\infty_\varepsilon((\nu_0, \Omega_0)) = \Big\{ (\nu, \Omega) \in \Theta_\nu \times \Theta_\Omega \enspace : \enspace \vert \nu - \nu_0 \vert < \varepsilon, \enspace d_\infty(\Omega, \Omega_0) < \varepsilon \Big\},
\]
where $d_\infty(A,B) = \max_{1\leq i,j\leq p} \vert a_{ij} - b_{ij}\vert$ denotes a max-norm distance for two $p\times p$ matrices. If $\nu_0>2$ and $\Omega_0$ is any $p\times p$ covariance matrix, then $\Pi\{ B^\infty_\varepsilon((\nu_0, \Omega_0)) \} > 0 $ for any $\varepsilon > 0$.}
\end{theorem}
\begin{theorem}
\label{thm:2}
 \textit{For fixed $\nu_0$ and $\Omega_0$, and for any $\varepsilon >0$, there exists $\varepsilon^\ast > 0 $, such that
\[
B^\infty_\varepsilon((\nu_0, \Omega_0))  \subset \Big\{ (\nu,\Omega)\in \Theta_\nu \times \Theta_\Omega : \text{KL}\big((\nu_0,\Omega_0), (\nu,\Omega)\big) < \varepsilon \Big\},
\]
where $\text{KL}((\nu_0,\Omega_0), (\nu,\Omega))$ denotes the Kullback-Leibler divergence between two multivariate Student-t distribution, $ t(\nu_0,\mathbf 0, \Omega_0)$ and $t(\nu,\mathbf 0, \Omega)$.}
\end{theorem}
Theorem~\ref{thm:1} states that the support of prior $\Pi$ is large enough so that arbitrarily small neighborhood of any $(\nu_0, \Omega_0) \in \Theta_\nu \times \Theta$ has strictly positive prior probability. Along with Theorem~\ref{thm:1}, Theorem~\ref{thm:2} ensures that, Kullback-Leibler support condition is satisfied for any $(\nu, \Omega)$ for the proposed prior $\Pi$. Thus if we prespecify $t$ degrees of freedom correctly, i.e., if we choose $\Pi_\nu = \delta_{\nu_0}$ for true $t$ degrees of freedom $\nu_0$, the weak posterior consistency holds by \citet{schwartz1965bayes}.

\section{Simulation Study}
\label{sec:4}
In this section, we illustrate a simulation study of covariance estimation under high-dimensional data and compare its performance with the normal-likelihood factor model of \citet{bhattacharya2011sparse}. We generated $\mathbf y_i, i=1, \ldots, n$ from heavy-tailed multivariate $t$ distribution with parameter $\nu_0$ and $\Omega_0 = \Lambda_0 \Lambda_0^T + \Sigma_0$. The true covariance of synthetic data is then $\frac{\nu_0}{\nu_0 -2} \Omega_0$. We let factor loading $\Lambda_0$ be sparse so that $70$--$80\%$ of entries of $\Omega_0$ are zero. The diagonal terms of error variance matrix $\Sigma_0$ is generated by the inverse gamma distribution of shape $1$ and rate $1/4$. Code for estimating covariance using the proposed model is available on \href{https://github.com/lee-jaejoon/robust-sparse-bayesian-infinite-factor-models}{https://github.com/lee-jaejoon/robust-sparse-bayesian-infinite-factor-models}.

The covariance estimation is conducted in two cases: when $\nu$ is well-specified and misspecified. In the well-specified case, the true degrees of freedom $\nu_0$ and the prespecified degrees of freedom $\nu$ in the model were set as $\nu_0 = \nu = 3$. In misspecified case, the degrees of freedom was $\nu=3$, while the true degrees of freedom was $\nu_0 = 7$. For each settings of $(p,k)$, 10 repeated simulations were conducted. We ran 20,000 iterations of Markov Chain Monte Carlo as described in Sect.~\ref{sec:3.2} with 5,000 burn-in steps. Learning rate $\epsilon$ for updating $\eta$ is set at $\epsilon = 0.025, 0.015, 0.01$ for $(p,k)=(200,10)$, $(500,15)$, $(1000, 20)$, respectively.  The adaptation probability in $t$ th iteration $p(t)$ is chosen $p(t) = \exp(-1.2 - 0.0004t)$. In the adaptation step, we deleted the factors $70\%$ of whose loading entries are closer to $0$ than $0.01$. The proposal variances of Metropolis-Hastings update for $a_1$ and $a_2$ are tuned so that the acceptance rates be $50$--$70\%$. After sampling from the posterior distribution is done, the covariance estimate is obtained by averaging the posterior samples of covariance. The estimated covariance is then evaluated with the matrix 1-norm (maximum absolute column sum), the matrix 2-norm (maximum singular value), the mean squared error (MSE), the average absolute bias (AAB), and the maximum absolute bias (MAB). The simulation result for well-specified case and misspecified case are displayed in Table~\ref{tab:1} and \ref{tab:2}, respectively.
\begin{table}
\setlength\tabcolsep{0pt}
\renewcommand{\arraystretch}{1.2}
{
\footnotesize
\begin{tabular*}{\textwidth}{@{\extracolsep{\fill}} c c c c c c c c c c c c}
\toprule
\multicolumn{2}{c}{\textbf{Model}} &
\multicolumn{5}{c}{$\quad$\textbf{Normal likelihood}$\quad$} & 
\multicolumn{5}{c}{\textbf{Multivariate }$t$ \textbf{likelihood}} \\
\cmidrule(l){3-7} \cmidrule(l){8-12}  
$(p,k)$ & & 1-norm & 2-norm & MSE & AAB & MAB & 1-norm & 2-norm & MSE & AAB & MAB  \\
\midrule 
& mean
& 33.3858 & 9.1841 & 0.0098 & 0.0561 & 1.0011 
& 30.0996 & 8.1040 & 0.0083 & 0.0530 & 0.8983\\
$\enspace p = 200$ & min
& 29.3328 & 8.1465 & 0.0077 & 0.0496 & 0.9280 
& 28.7672 & 8.0434 & 0.0071 & 0.0486 & 0.8428\\
$\enspace k = 10$ & median
& 33.4496 & 9.0973 & 0.0099 & 0.0566 & 0.9881 
& 29.8370 & 8.1069 & 0.0081 & 0.0526 & 0.9042\\
& max
& 36.4696 & 11.3066 & 0.0111 & 0.0593 & 1.0978 
& 32.1881 & 8.1538 & 0.0093 & 0.0562 & 0.9511\\
\midrule 
& mean
& 89.6723 & 25.3974 & 0.0116 & 0.0677 & 1.0639 
& 78.2845 & 23.2956 & 0.0100 & 0.0656 & 0.9032\\
$\enspace p = 500$ & min
& 82.4762 & 23.7311 & 0.0101 & 0.0638 & 0.9714 
& 74.6951 & 23.1118 & 0.0088 & 0.0609 & 0.8452\\
$\enspace k = 15$ & median
& 85.1119 & 24.2223 & 0.0118 & 0.0684 & 1.0143 
& 78.1479 & 23.2990 & 0.0098 & 0.0649 & 0.8963\\
& max
& 107.7094 & 31.6338 & 0.0127 & 0.0715 & 1.2807 
& 82.4753 & 23.5229 & 0.0117 & 0.0714 & 0.9890\\
\midrule 
& mean
& 217.1357 & 46.8709 & 0.0142 & 0.0752 & 1.6856 
& 205.3249 & 37.6516 & 0.0131 & 0.0755 & 1.4546\\
$\enspace p = 1000$ & min
& 198.5997 & 38.9138 & 0.0129 & 0.0715 & 1.5203 
& 200.2763 & 37.5020 & 0.0116 & 0.0716 & 1.3352\\
$\enspace k = 20$ & median
& 217.1729 & 45.1398 & 0.0132 & 0.0735 & 1.6032 
& 205.5342 & 37.6245 & 0.0134 & 0.0763 & 1.4816\\
& max
& 234.4568 & 57.5753 & 0.0164 & 0.0809 & 1.9884 
& 211.1231 & 37.8733 & 0.0141 & 0.0786 & 1.5669\\
\bottomrule    
\end{tabular*}
}
\caption{The simulation result of the covariance estimation when the true degrees of freedom is $\nu_0 = 3$ and the model degrees of freedom is $\nu=3$ \label{tab:1}}
\end{table}

\begin{table}
\setlength\tabcolsep{0pt}
\renewcommand{\arraystretch}{1.2}
{
\footnotesize
\begin{tabular*}{\textwidth}{@{\extracolsep{\fill}} c c c c c c c c c c c c}
\toprule
\multicolumn{2}{c}{\textbf{Model}} &
\multicolumn{5}{c}{$\quad$\textbf{Normal likelihood}$\quad$} & 
\multicolumn{5}{c}{\textbf{Multivariate }$t$ \textbf{likelihood}} \\
\cmidrule(l){3-7} \cmidrule(l){8-12}  
$(p,k)$ & & 1-norm & 2-norm & MSE & AAB & MAB & 1-norm & 2-norm & MSE & AAB & MAB  \\
\midrule 
& mean
& 36.7584 & 10.9573 & 0.0105 & 0.0573 & 1.4114 
& 30.9032 & 8.5313 & 0.0088 & 0.0544 & 1.3421\\
$\enspace p = 200$ & min
& 30.5850 & 8.5565 & 0.0085 & 0.0511 & 1.2321 
& 25.9585 & 7.1489 & 0.0078 & 0.0517 & 1.1556\\
$\enspace k = 10$ & median
& 35.1613 & 10.6633 & 0.0100 & 0.0561 & 1.3626 
& 31.4622 & 8.5725 & 0.0087 & 0.0546 & 1.3700\\
& max
& 57.0237 & 17.8385 & 0.0178 & 0.0766 & 1.6724 
& 36.5888 & 10.1715 & 0.0104 & 0.0587 & 1.4878\\
\midrule 
& mean
& 91.8864 & 24.8418 & 0.0123 & 0.0709 & 1.1925 
& 81.9975 & 24.4893 & 0.0090 & 0.0617 & 1.0973\\
$\enspace p = 500$ & min
& 81.3002 & 24.5005 & 0.0102 & 0.0647 & 1.0277 
& 79.2123 & 24.4243 & 0.0083 & 0.0587 & 0.9954\\
$\enspace k = 15$ & median
& 90.7399 & 24.5695 & 0.0117 & 0.0699 & 1.1620 
& 81.3976 & 24.4942 & 0.0088 & 0.0608 & 1.0718\\
& max
& 104.1514 & 25.6546 & 0.0150 & 0.0791 & 1.5117 
& 86.9735 & 24.5355 & 0.0100 & 0.0663 & 1.2767\\
\midrule 
& mean
& 196.8423 & 44.7915 & 0.0113 & 0.0655 & 1.6796
& 193.8782 & 39.5985 & 0.0137 & 0.0769 & 1.3940\\
$\enspace p = 1000$ & min
& 177.2834 & 39.5988 & 0.0108 & 0.0647 & 1.5423
& 188.0602 & 39.3843 & 0.0130 & 0.0746 & 1.3285\\
$\enspace k = 20$ & median
& 200.2282 & 43.2991 & 0.0114 & 0.0652 & 1.6653
& 193.3312 & 39.4151 & 0.0133 & 0.0764 & 1.3789\\
& max
& 213.9246 & 55.1276 & 0.0118 & 0.0667 & 1.9318
& 200.3391 & 40.3553 & 0.0151 & 0.0809 & 1.4485\\
\bottomrule    
\end{tabular*}
}
\caption{The simulation result of the covariance estimation when the true degrees of freedom is $\nu_0 = 7$ and the model degrees of freedom is $\nu=3$ \label{tab:2}}
\end{table}
Table~\ref{tab:1} shows the simulation results of the well-specified case where both true and model degrees of freedoms are $\nu_0=\nu = 3$ for covariance estimation. The proposed model performs better than the normal likelihood model in all cases. In $(p,k)=(1000,20)$, MSE and AAB of normal likelihood model shows smaller value than that of the proposed $t$ likelihood model. However, observing that maximum absolute bias of normal likelihood model is larger, we can presume that the scale of covariance entries is underestimated in normal likelihood model's case, which leads to biased estimation. Though the estimation performance was slightly poor, the normal-likelihood factor model estimated the number of factors in a stable manner, even with the data from heavy-tailed distribution. Mean elapsed times for the proposed model are $4.18$, $15.51$, $46.19$ minutes, which are about $1.52$, $1.47$, $1.50$ times longer than those of the normal model in $(p,k)=(200,10)$, $(500,15)$, $(1000, 20)$, respectively. As we set up sparse true covariance matrix 70 to 80\% of whose entries are zero, we can monitor and compare the covariance estimates of the two model for those strictly zero covariance entries. For covariance entries whose true values are zero, 10th and 90th percentile of estimated covariance entries from the proposed model are $(-0.0608, 0.0816)$, $(-0.0873, 0.1010)$, $(-0.0992, 0.1097)$ on average, while the normal model showed $(-0.0651, 0.0813)$, $(-0.0835, 0.0937)$, $(-0.0950, 0.1003)$ in $(p,k)=(200,10)$, $(500,15)$, $(1000, 20)$, respectively. This demonstrates that the proposed model and the normal model have similar shrinkage for the true zero entries.

Table~\ref{tab:2} shows the simulation results of the misspecified case where model degrees of freedom is $\nu = 3$ while true degrees of freedom is $\nu_0=7$. Even when the degrees of freedom is misspecified, we can see that using the proposed model with small enough degrees of freedom yields better covariance estimation performance than the normal model. Likewise, we can observe the possible bias in normal likelihood model when $(p,k) = (1000,20)$ as in Table~\ref{tab:1}. Also the proposed model does not lose the capability of estimating the number of latent factors under misspecification of the degrees of freedom. Mean elapsed times for the proposed model are $4.43$, $17.19$, $48.86$ minutes, which are about $1.60$, $1.59$, $1.58$ times longer than those of the normal model in $(p,k)=(200,10)$, $(500,15)$, $(1000, 20)$, respectively. From the proposed model, the 10th and 90th percentile of estimated covariance entries whose true values are zero are $(-0.0612, 0.0892)$, $(-0.0744, 0.0890)$, $(-0.1023, 0.1133)$ on average, while the normal model showed $(-0.0640, 0.0858)$, $(-0.0882, 0.1023)$, $(-0.0791, 0.0860)$ in $(p,k)=(200,10)$, $(500,15)$, $(1000, 20)$, respectively. This implies that, even under misspecified degrees of freedom, the proposed model still shows similar shrinkage for true zero entries compared to normal model.

\section{Real Data Analysis : T1T2 Node-Negative Breast Cancer Application}
\label{sec:5}
\subsection{Background and Previous Researches}
\label{sec:5.1}
Carcinoma is a type of cancer that develops from epithelial cells. Invasive ductal carcinoma is a type of breast carcinoma which begins growing in a milk duct and invades adjacent tissue of the breast. It is the most common type of breast cancer, accounting for 80\% of all breast cancer diagnoses. Cancer cells are developed by accumulations of multiple DNA mutations that are not repaired by their own repair mechanisms. \citet{gravier2010prognostic} analyzed the DNA signature of tumor cells from 168 patients with small invasive ductal carcinomas without axillary lymph node involvement (T1T2N0) to predict metastasic progression in 5 years after diagnosis.

\begin{figure}
\centering
\includegraphics[width=0.75\textwidth]{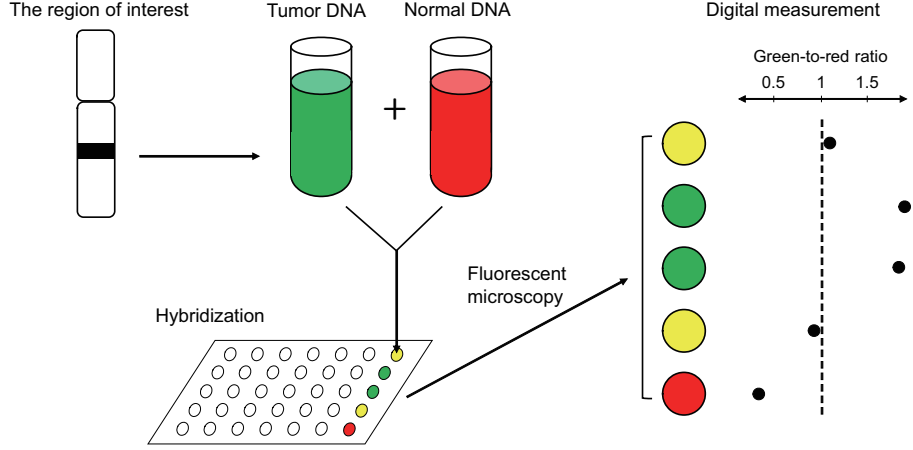}
\caption{The overview of array comparative genomic hybridization (aCGH) procedure}
\label{fig:2}
\end{figure}

Gene expression of each patient's tumor cell was obtained by array comparative genomic hybridization(aCGH). aCGH is a technique to detect the change in chromosomal copy number. DNAs of tumor cell and normal cell are labelled with green and red fluorescent protein, respectively. The DNAs are then mixed and undergone hybridization: the process of single stranded DNA binding to its complementary DNA strand. Next, green-to-red ratio is measured by fluorescent microscopy, which represents the chromosomal gain or loss of tumor DNA in the region of interest. The overall procedure of data acquisition through aCGH is illustrated in Fig.~\ref{fig:2}. 

The training set contained 2,905 predictor variables (log2 transformed) representing genomic signatures of chromosome 2p22.2, 3p23, and 8q21-24. Among 168 patients, 111 patients did not have any metastatic event in 5 years after initial diagnosis, while early metastasis of breast carcinoma was reported in other 57 patients. The dataset analysed during the current study is available in the Gene Expression Omnibus (GEO) repository database with accession number GSE19159, \href{https://www.ncbi.nlm.nih.gov/geo/query/acc.cgi?acc=GSE19159}{https://www.ncbi.nlm.nih.gov/geo/query/acc.cgi?acc=GSE19159}. In order to predict the progression of metastasis, \citet{gravier2010prognostic} combined the outcome of multiple classifiers each of which are based on logistic regression.

The latent factor regression is an efficient method under high-dimensional setting of $p \gg n$, where joint covariance structure of continuous dependent variable $z_i$ and predictor variable $\mathbf x_i$ is estimated by performing factor model on $\mathbf y_i = (z_i, \mathbf x_i^T)^T$. The predictive distribution for $z_\text{new}$ can be obtained as follows:
\begin{equation*}
p(z_\text{new} \vert \mathbf x_\text{new}, \mathbf y_1, \ldots , \mathbf y_n) = \int p(z_\text{new} \vert \mathbf x_\text{new}, \Omega) p(\Omega \vert y_1, \ldots , \mathbf y_n) d\Omega,
\end{equation*}
Under joint normality assumption, the conditional distribution of $z$ given $\mathbf x, \Omega$ is $\mathcal N ( \mathbf x^T \Omega_{xx}^{-1}\Omega_{xz}, \Omega_{zz} - \Omega_{zx}\Omega_{xx}^{-1}\Omega_{xz})$. Here $\beta = \Omega_{xx}^{-1}\Omega_{xz}$ can be considered as a regression coefficient in latent factor regression. There have been a few approaches to analyze high-dimensional microarray data with latent factor regression \citep{carvalho2008high, bhattacharya2011sparse}. \citet{bhattacharya2011sparse} implemented their shrinkage prior to induce shrinkage in regression coefficient estimate. Then the feature selection is performed by sorting predictor variables by absolute value of estimated regression coefficient.

\subsection{Model and Results}
\label{sec:5.2}
Our goal is to predict the progression of metastasis given DNA signature data of cancer cells (0: no metastasis, 1: metastasis). By investigating Q-Q plot of each variable, we have observed a heavy-tailed structure of the data. Though the proposed factor model is an efficient tool to estimate the low-dimensional structure of heavy-tailed high-dimensional data, we cannot implement the latent factor regression method with the proposed model because the dependent variable is not continuous but binary.

Instead, we implemented discriminant analysis using the covariance estimate obtained by the proposed model. We divide the data set into the training set ($118$ of $168$ patients) and the test set ($50$ of $168$ patients). Covariance estimates for patients with metastasis ($36$ of $118$ patients) and without metastasis (82 of 118 patients) are obtained separately from the training set. We ran Markov chain Monte Carlo algorithm as in Sect.~\ref{sec:3.2} for posterior computation for $20,000$ iterations with $5,000$ burn-in steps. The degrees of freedom of $t$ likelihood is set to $\nu = 5$. The step size $\epsilon$ for Hamiltonian Monte Carlo update for $\eta_i, i = 1, \ldots, n$ is set to $\epsilon = 0.2$. The estimated number of factors for patients with metastasis and without metastasis are $50$ and $70$ with 95\% credible interval $(48,52)$ and $(65,71)$, respectively.

After estimating covariance for patients with metastasis and without metastasis, we calculated the log likelihood ratio of observations in test set as follows:
\begin{equation*}
\log\left(\frac{t(\mathbf y; \nu, \hat \mu_1, \hat \Sigma_1)}{t(\mathbf y; \nu, \hat \mu_0, \hat \Sigma_0)} \right) = \log\left(\frac{\left\vert \hat \Sigma_1 \right\vert^{-\frac{1}{2}} \left\{ \nu + (\mathbf y - \hat \mu_1)^T  \hat \Sigma_1 (\mathbf y - \hat \mu_1) \right\}^{-\frac{\nu + p}{2}}}{\left\vert \hat \Sigma_0 \right\vert^{-\frac{1}{2}} \left\{ \nu + (\mathbf y - \hat \mu_0)^T  \hat \Sigma_0 (\mathbf y - \hat \mu_0) \right\}^{-\frac{\nu + p}{2}}} \right), 
\end{equation*}
where $\hat \mu_1, \hat \mu_0$ are training sample mean of patients with metastasis and without metastasis, respectively. The covariance estimates of patients with metastasis and without metastasis obtained by the proposed model are denoted as $\hat \Sigma_1$ and $\hat \Sigma_0$, respectively. If the log likelihood ratio is greater than a threshold $\xi$, we classified the observation as a patient with metastasis. We determined the value of threshold $\xi = 0$ in our case. Sensitivity is the proportion of true positives which are correctly identified by classifier, while specificity is the proportion of true negatives which are correctly identified by classifier. Both are measures of classification performance widely used in medicine. The test accuracy was $86\%$ which outperforms the classfier suggested by \citet{gravier2010prognostic}. The classfier of \citet{gravier2010prognostic} showed test accuracy of $78\%$. Test sensitivity of $66.7\%$ and test specificity of $90.2\%$ are observed, while \citet{gravier2010prognostic} showed $84\%$ and $66\%$, respectively.

\section{Discussion}
\label{sec:6}
In this paper, we have proposed a Bayesian infinite factor model with multiplicative gamma process shrinkage prior for robust covariance estimation under heavy-tailed high-dimensional data. Also we have shown the fact that, under well-specified degrees of freedom of $t$ distribution, the posterior density from the proposed model is weakly consistent.

There are a few research directions which are worthy of further study. \citet{kleijn2006misspecification} and \citet{ramamoorthi2015posterior} have studied posterior consistency under model misspecification. In the same spirit, theoretical properties of the proposed model under misspecification of the degrees of freedom can be potential avenues of exploration. \citet{murphy2020infinite} has introduced \textit{the infinite mixture of infinite factor analysers} (IMIFA) model, which is a Pitman-Yor mixture of the model of \citet{bhattacharya2011sparse}. The same extension of the proposed model from normal likelihood to Student's $t$-likelihood can also be made when some or all of the mixture components are suspected to follow heavy-tailed distribution. Finally, the proposed model is not completely choice-free, due to step size parameter $\epsilon$ used in No-U-Turn sampler update for $\eta$. \citet{hoffman2014no} suggested a method of adaptive setting for the value of $\epsilon$. This, however, is not directly applicable in our settings, because we are using a single iteration of No-U-Turn sampler whose target function changes as estimates of the other parameters change. Devising a method of tuning $\epsilon$ would be an improvement on our work.

\section*{Appendix}
\label{sec:7}
\subsection*{Proof of Theorem 1}
\label{sec:7.1}
Let $\varepsilon >0$ be fixed, and let 
\begin{equation*}
B_{\varepsilon}\big((\nu_0, \Lambda_0, \Sigma_0)\big) = \Big\{ (\nu, \Lambda, \Sigma) : \vert \nu - \nu_0 \vert < \varepsilon ,d_2( \Lambda, \Lambda_0 ) < \varepsilon,d_\infty( \Sigma , \Sigma_0 ) < \varepsilon\Big\}.
\end{equation*}
By Lemma 2 of \citet{bhattacharya2011sparse}, there exists $\varepsilon_1 > 0$ such that
\begin{equation*}
\begin{split}
\tilde g\Big(B_{\varepsilon_1}\big((\nu_0, \Lambda_0, \Sigma_0)\big) \Big) &\subset B_\varepsilon^\infty\Big( \tilde g\big((\nu_0,\Lambda_0, \Sigma_0) \big)\Big) \\
&= B_\varepsilon^\infty\big((\nu_0,g(\Lambda_0, \Sigma_0))\big) = B_\varepsilon^\infty\big((\nu_0,\Omega_0)\big).  
\end{split}
\end{equation*}
Thus, we have 
\begin{equation*}
\begin{split}
B_{\varepsilon_1}\big((\nu_0, \Lambda_0, \Sigma_0)\big)  &\subset \tilde g^{-1}\Big( B_\varepsilon^\infty\big((\nu_0,\Omega_0)\big) \Big).
\end{split}
\end{equation*}
Denoting the prior distribution as $\Pi = \Pi_\nu \otimes \Pi_\Omega =(\Pi_\nu \otimes \Pi_\Lambda \otimes \Pi_\Sigma)\circ \tilde g^{-1}$, we have
\begin{equation*}
\begin{split}
(\Pi_\nu \otimes \Pi_\Lambda \otimes \Pi_\Sigma)\Big\{B_{\varepsilon_1}\big((\nu_0, \Lambda_0, \Sigma_0)\big)\Big\}  &\leq  (\Pi_\nu \otimes \Pi_\Lambda \otimes \Pi_\Sigma)\Big\{ \tilde g^{-1}\Big( B_\varepsilon^\infty\big((\nu_0,\Omega_0)\big) \Big) \Big\} \\
&= \Pi\Big(B_\varepsilon^\infty\big((\nu_0,\Omega_0)\big)\Big).
\end{split}
\end{equation*}
Thus, if 
\begin{equation*}
\begin{split}
    &\Pi_\nu\Big( \big\{ \nu \in \Theta_\nu: \vert \nu - \nu_0 \vert < \varepsilon_1 \big\}\Big) > 0 \\
    &\Pi_\Lambda\Big( \big\{ \Lambda \in \Theta_\Lambda: d_2( \Lambda, \Lambda_0 ) < \varepsilon_1 \big\}\Big) > 0 \\
    &\Pi_\Sigma\Big( \big\{ \Sigma \in \Theta_\Sigma: d_\infty( \Sigma , \Sigma_0 ) < \varepsilon_1 \big\}\Big) > 0, 
\end{split}
\end{equation*}
we obtain the conclusion.

Since the support of $\Pi_\nu$ and $\Pi_\Sigma$ are $\Theta_\nu$ and $\Theta_\Sigma$, respectively,  the inequalities for $\nu$ and $\Sigma$ hold.  For the inequality of the $\Lambda$,  we can apply the proof of Proposition 2 of \citet{bhattacharya2011sparse}. \qed

\subsection*{Proof of Theorem 2}
\label{sec:7.2}
Let $\nu_0 > 2, \Omega_0 \in \Theta_\Omega$ be true parameter. We wish to show that, for any $\varepsilon > 0$, we can choose $\varepsilon^\ast > 0$ such that
\begin{equation}
\text{KL}\Big((\nu_0,\Omega_0), (\nu,\Omega) \Big) < \varepsilon, ~ \text{ for all } \vert\nu_0-\nu \vert < \varepsilon^\ast \text{ and }d_\infty(\Omega_0, \Omega) < \varepsilon^\ast. \label{WTS}
\end{equation}
Let $\varepsilon > 0 $ be given. By the definition of Kullback-Leibler divergence, we have
\begin{equation}
\begin{split}
&\text{KL}\Big((\nu_0,\Omega_0), (\nu,\Omega) \Big) \\
&= \int \log \frac{t(\mathbf y; \nu_0, \Omega_0)}{t(\mathbf y; \nu,\Omega)}t(\mathbf y; \nu_0, \Omega_0)d\mathbf y \\
&= \mathbb E_{(\nu_0, \Omega_0)}\left[ \log \left( \frac{\frac{\Gamma[(\nu_0 + p)/2 ]}{\Gamma(\nu_0/2)(\nu_0\pi)^{p/2} \det(\Omega_0)^{1/2}} \left[ 1 + \frac{\mathbf y^T \Omega_0^{-1}\mathbf y}{\nu_0}\right]^{-{(\nu_0 + p)}/{2}}}{\frac{\Gamma[(\nu + p)/2 ]}{\Gamma(\nu/2)(\nu\pi)^{p/2} \det(\Omega)^{1/2}} \left[ 1 + \frac{\mathbf y^T \Omega^{-1}\mathbf y}{\nu}\right]^{-{(\nu + p)}/{2}}}\right)\right] \\
&= \log\left( \frac{\frac{\Gamma[(\nu_0 + p)/2 ]}{\Gamma(\nu_0/2) }\nu_0^{\nu_0/2}}{\frac{\Gamma[(\nu + p)/2 ]}{\Gamma(\nu/2)}\nu^{\nu/2}} \right)+ \frac{1}{2}\log \left( \frac{\det(\Omega)}{\det(\Omega_0)}\right) + \mathbb E_{(\nu_0, \Omega_0)}\left[ \log \frac{\left[ \nu_0 + \mathbf y^T \Omega_0^{-1}\mathbf y\right]^{-{(\nu_0 + p)}/{2}}}{\left[ \nu + \mathbf y^T \Omega^{-1}\mathbf y\right]^{-{(\nu + p)}/{2}}} \right] \\
&\leq \left\vert \log\left( \frac{\Gamma[(\nu_0 + p)/2 ]}{\Gamma(\nu_0/2) }\nu_0^{\nu_0/2} \right) - \log\left( {\frac{\Gamma[(\nu + p)/2 ]}{\Gamma(\nu/2)}\nu^{\nu/2}} \right) \right\vert \\
&\enspace \enspace + \left\vert\frac{1}{2} \log \left( \det(\Omega)\right) - \frac{1}{2}\log\left(\det(\Omega_0)\right) \right\vert + \left\vert \mathbb E_{(\nu_0, \Omega_0)}\left[ \log \frac{\left[ \nu_0 + \mathbf y^T \Omega_0^{-1}\mathbf y\right]^{-\frac{\nu_0 + p}{2}}}{\left[ \nu + \mathbf y^T \Omega^{-1}\mathbf y\right]^{-\frac{\nu + p}{2}}} \right]\right\vert. 
\end{split}\label{KLupper}
\end{equation}
By continuity of the functions in the Eq.~\ref{KLupper}, we can choose $\varepsilon^\ast_1$, $\varepsilon^\ast_2 >0$ that bounds the first and second terms of Eq.~\ref{KLupper} with $\varepsilon/3$, respectively. By the triangle inequality, the third term of Eq.~\ref{KLupper} is
\begin{equation}
\begin{split}
&\left\vert \mathbb E_{(\nu_0, \Omega_0)}\left[ \log \frac{\left[ \nu_0 + \mathbf y^T \Omega_0^{-1}\mathbf y\right]^{-{(\nu_0 + p)}/{2}}}{\left[ \nu + \mathbf y^T \Omega^{-1}\mathbf y\right]^{-{(\nu + p)}/{2}}} \right]\right\vert \\
&\leq \left\vert \frac{\nu + p}{2} \mathbb E\log \left[ \nu + \mathbf y^T \Omega^{-1}\mathbf y\right] - \frac{\nu_0 + p}{2} \mathbb E\log \left[ \nu + \mathbf y^T \Omega^{-1}\mathbf y\right] \right\vert \\
&\enspace + \left\vert \frac{\nu_0 + p}{2} \mathbb E\log \left[ \nu + \mathbf y^T \Omega^{-1}\mathbf y\right] - \frac{\nu_0 + p}{2} \mathbb E\log \left[ \nu_0 + \mathbf y^T \Omega_0^{-1}\mathbf y\right] \right\vert \label{AplusB}\\
&= A + B .
\end{split}
\end{equation}
Denote the first and second terms of Eq.~\ref{AplusB} as $A$ and $B$, respectively. For $A$, we have
\begin{equation*}
\begin{split}
A &= \frac{\vert \nu -\nu_0 \vert}{2} \left\vert \mathbb E\log \left[ \nu + \mathbf y^T \Omega^{-1}\mathbf y\right] \right\vert \\
&\leq \frac{\vert \nu -\nu_0 \vert}{2}  \mathbb E \Big[ \left\vert \log \left[ \nu + \mathbf y^T \Omega^{-1}\mathbf y\right] \right\vert \Big]\\
&= \frac{\vert \nu -\nu_0 \vert}{2}  \mathbb E \Big[ \log \left[ \nu + \mathbf y^T \Omega^{-1}\mathbf y\right]  \Big] \\
&\leq \frac{\vert \nu -\nu_0 \vert}{2}  \mathbb E \Big[ \nu-1 + \mathbf y^T \Omega^{-1}\mathbf y \Big] .
\end{split}
\end{equation*}
Using the fact that the expectation of quadratic form of $\mathbf y \sim t(\nu_0, \mathbf 0, \Omega_0)$ is $\mathbb E[\mathbf y^T \Omega^{-1}\mathbf y] = \frac{\nu_0}{\nu_0-2}\text{tr}(\Omega^{-1} \Omega_0)$, we have
\begin{equation*}
\begin{split}
A &= \frac{\vert \nu -\nu_0 \vert}{2} \Big[ \nu-1 + \frac{\nu_0}{\nu_0-2}\text{tr}(\Omega^{-1} \Omega_0)\Big] \\
&= \frac{\vert \nu -\nu_0 \vert}{2} \left[ \nu-1 + \frac{\nu_0}{\nu_0-2}\sum_{j=1}^{p}\lambda_{j}(\Omega^{-1} \Omega_0) \right] \\
&\leq \frac{\vert \nu -\nu_0 \vert}{2} \Big[ \vert \nu -\nu_0 \vert + \nu_0 -1 + \frac{\nu_0}{\nu_0-2}p\lambda_{\max}(\Omega^{-1} \Omega_0)\Big].
\end{split}
\end{equation*}
Let $\lambda_{\max}(\Omega^{-1} \Omega_0)$ be the largest eigenvalue of $\Omega^{-1} \Omega_0$. For an eigenvector $v \in \mathbb R^p$ corresponding to $\lambda_{\max}(\Omega^{-1} \Omega_0)$ and sufficiently large $M_1 > 0$, the following holds:
\begin{equation*}
\begin{split}
\lambda_{\max}(\Omega^{-1} \Omega_0) &= \Vert \lambda_{\max}(\Omega^{-1} \Omega_0) v \Vert_2 \\
&\leq p^{1/2} \Vert \lambda_{\max}(\Omega^{-1} \Omega_0) v \Vert_\infty \\
&= p^{1/2} \Vert\Omega^{-1} \Omega_0 v \Vert_\infty \\
&\leq p^{1/2} \Vert\Omega^{-1} \Vert_\infty \Vert \Omega_0 \Vert_\infty \Vert v \Vert_\infty \\ 
&\leq  p^{1/2} (\Vert\Omega^{-1} - \Omega_0^{-1} \Vert_\infty + \Vert \Omega_0^{-1}\Vert_\infty) \Vert \Omega_0 \Vert_\infty \Vert v \Vert_\infty \\
&\leq  p^{1/2} (\Vert\Omega^{-1} - \Omega_0^{-1} \Vert_\infty + M_1) M_1 \Vert v \Vert_\infty \\
&\leq  p^{1/2} (\Vert\Omega^{-1} - \Omega_0^{-1} \Vert_\infty + M_1) M_1 \\
&=  p^{1/2} (\Vert\Omega^{-1} - \Omega_0^{-1} \Vert_\infty + M_1) M_1 .
\end{split}
\end{equation*}
With this upper bound of $\lambda_{\max}(\Omega^{-1} \Omega_0)$, we have
\begin{equation*}
\begin{split}
A &\leq \frac{\vert \nu -\nu_0 \vert}{2} \Big[ \vert \nu -\nu_0 \vert + \nu_0 -1 + \frac{\nu_0}{\nu_0-2}p\lambda_{\max}(\Omega^{-1} \Omega_0)\Big] \\
&\leq \frac{\vert \nu -\nu_0 \vert}{2} \Big[ \vert \nu -\nu_0 \vert + \nu_0 -1 + \frac{\nu_0}{\nu_0-2}p^{3/2} (\Vert\Omega^{-1} - \Omega_0^{-1} \Vert_\infty + M_1) M_1\Big].
\end{split}
\end{equation*}
By continuity of matrix inversion, we can choose $\tilde \varepsilon >0$ such that $\Vert\Omega - \Omega_0 \Vert_\infty < \tilde \varepsilon $  implies $\Vert\Omega^{-1} - \Omega_0^{-1} \Vert_\infty < 1$. Plus we can choose $\varepsilon^\ast_3 \in (0, \tilde \varepsilon)$ small enough so that $A$ is bounded above by $\varepsilon / 6$. So we have
\begin{equation*}
\begin{split}
A &< \frac{\vert \nu -\nu_0 \vert}{2} \Big[ \vert \nu -\nu_0 \vert + \nu_0 -1 + \frac{\nu_0}{\nu_0-2}p^{3/2} (1 + M_1) M_1\Big] \\
&< \frac{\varepsilon^\ast_3}{2} \Big[ \varepsilon^\ast_3 + \nu_0 -1 + \frac{\nu_0}{\nu_0-2}p^{3/2} (1 + M_1) M_1\Big] \\
&< \frac{\varepsilon}{6}.
\end{split}
\end{equation*}
For $B$, by Jensen's inequality, we have
\begin{equation}
\begin{split}
B &= \left\vert \frac{\nu_0 + p}{2} \mathbb E\log \left[ \nu + \mathbf y^T \Omega^{-1}\mathbf y\right] - \frac{\nu_0 + p}{2} \mathbb E\log \left[ \nu_0 + \mathbf y^T \Omega_0^{-1}\mathbf y\right] \right\vert  \\
&=\frac{\nu_0 + p}{2}  \left\vert \mathbb E\log \left[ \frac{\nu + \mathbf y^T \Omega^{-1}\mathbf y}{\nu_0 + \mathbf y^T \Omega_0^{-1}\mathbf y}\right] \right\vert\\
&\leq \frac{\nu_0 + p}{2} \mathbb E \left\vert \log \left[ \frac{\nu + \mathbf y^T \Omega^{-1}\mathbf y}{\nu_0 + \mathbf y^T \Omega_0^{-1}\mathbf y}\right] \right\vert. \label{Bbound}
\end{split}
\end{equation}
For a fixed unit vector $\omega \in \mathbb R^p$, let $g_\omega(t)$ be a function defined on $t>0$ as follows:
\begin{equation*}
\begin{split}
g_\omega(t) &= \frac{\nu + \mathbf y^T \Omega^{-1}\mathbf y}{\nu_0 + \mathbf y^T \Omega_0^{-1}\mathbf y} \Bigg\vert_{\mathbf y = t \omega} \\
&= \frac{\nu + t^2 \omega^T \Omega^{-1}\omega}{\nu_0 + t^2\omega^T \Omega_0^{-1}\omega}.
\end{split}
\end{equation*}
Investigating critical points and limits of $t>0$, we have the following bound of $g_\omega(t)$,
\begin{equation}
\begin{split}
\frac{\omega^T \Omega^{-1}\omega}{\omega^T \Omega_0^{-1}\omega} \wedge \frac{\nu}{\nu_0} \leq g_\omega(t) \leq \frac{\omega^T \Omega^{-1}\omega}{\omega^T \Omega_0^{-1}\omega} \vee\frac{\nu}{\nu_0}. \label{gomegatsolution}
\end{split}
\end{equation}
Eq.~\ref{gomegatsolution} holds for any unit vector $\omega \in \mathbb R^p$. Thus by taking infimum and supremum on lower and upper bounds, respectively, we have
\begin{equation}
\begin{split}
\left[ \left(\inf_{\Vert \omega \Vert = 1}\frac{\omega^T \Omega^{-1}\omega}{\omega^T \Omega_0^{-1}\omega} \right) \wedge \frac{\nu}{\nu_0} \right] \leq g_\omega(t) \leq \left[ \left( \sup_{\Vert \omega \Vert = 1} \frac{\omega^T \Omega^{-1}\omega}{\omega^T \Omega_0^{-1}\omega} \right) \vee \frac{\nu}{\nu_0} \right]. \label{gomegatbound}
\end{split}
\end{equation}
For $\tilde \omega = {\Omega_0^{-1/2}\omega}/{\Vert \Omega_0^{-1/2}\omega\Vert_2}$, we yield the following inequality of $\frac{\omega^T \Omega^{-1}\omega}{\omega^T \Omega_0^{-1}\omega} $,
\begin{equation}
\begin{split}
\lambda_{\min} (\Omega_0^{1/2}\Omega^{-1}\Omega_0^{1/2}) \leq \frac{\omega^T \Omega^{-1}\omega}{\omega^T \Omega_0^{-1}\omega} \leq \lambda_{\max} (\Omega_0^{1/2}\Omega^{-1}\Omega_0^{1/2}), \label{boundbound}
\end{split}
\end{equation}
which is obtained by following result,
\begin{equation*}
\begin{split}
\frac{\omega^T \Omega^{-1}\omega}{\omega^T \Omega_0^{-1}\omega} &= \frac{\omega^T \Omega_0^{-1/2}\Omega_0^{1/2}\Omega^{-1}\Omega_0^{1/2}\Omega_0^{-1/2}\omega}{\omega^T \Omega_0^{-1/2}\Omega_0^{-1/2}\omega} \\
&= \frac{\tilde \omega^T \Omega_0^{1/2}\Omega^{-1}\Omega_0^{1/2}\tilde \omega}{\tilde \omega^T\tilde \omega}.
\end{split}
\end{equation*}
Here $\lambda_{\min}(\Omega_0^{1/2}\Omega^{-1}\Omega_0^{1/2})$ and $\lambda_{\max}(\Omega_0^{1/2}\Omega^{-1}\Omega_0^{1/2})$ are the smallest and the largest eigenvalues of $\Omega_0^{1/2}\Omega^{-1}\Omega_0^{1/2}$, respectively. By Eq.~\ref{gomegatbound} and Eq.~\ref{boundbound}, $\log g_\omega(t)$ is bounded as follows:
\begin{equation*}
\left[ \lambda_{\min}(\Omega_0^{1/2}\Omega^{-1}\Omega_0^{1/2}) \wedge \frac{\nu}{\nu_0} \right] \leq g_\omega(t) \leq \left[ \lambda_{\max}(\Omega_0^{1/2}\Omega^{-1}\Omega_0^{1/2}) \vee \frac{\nu}{\nu_0}, \right]
\end{equation*}
\begin{equation}
\log \left[ \lambda_{\min}(\Omega_0^{1/2}\Omega^{-1}\Omega_0^{1/2}) \wedge \frac{\nu}{\nu_0} \right] \leq \log g_\omega(t) \leq \log \left[ \lambda_{\max}(\Omega_0^{1/2}\Omega^{-1}\Omega_0^{1/2}) \vee \frac{\nu}{\nu_0} \right]. \label{loggomegatbound}
\end{equation}
Note that Eq.~\ref{loggomegatbound} holds for any $ \omega \in \mathbb R^p, \Vert \omega \Vert_2 = 1$. For any $\mathbf y \in \mathbb R^p$, $\mathbf y$ can be written as $\mathbf y = \Vert \mathbf y \Vert \frac{\mathbf y}{\Vert \mathbf y \Vert} = t \omega, \enspace t \stackrel{let}{=}\Vert \mathbf y \Vert,\enspace \omega \stackrel{let}{=} \frac{\mathbf y}{\Vert \mathbf y \Vert}$. Thus we have the upper bound of the integrand of Eq.~\ref{Bbound} as follows:
\begin{equation*}
\begin{split}
&\left\vert \log\left[\frac{\nu + \mathbf y^T \Omega^{-1}\mathbf y}{\nu_0 + \mathbf y^T \Omega_0^{-1}\mathbf y} \right] \right\vert \\
&\leq \max\left\{ \left\vert \log\left[ \lambda_{\min}(\Omega_0^{1/2}\Omega^{-1}\Omega_0^{1/2}) \wedge \frac{\nu}{\nu_0} \right] \right\vert , \left\vert \log\left[ \lambda_{\max}(\Omega_0^{1/2}\Omega^{-1}\Omega_0^{1/2}) \vee \frac{\nu}{\nu_0} \right] \right\vert \right\}.
\end{split}
\end{equation*}
Here we use the following limiting property of eigenvalue as $\Omega \to \Omega_0$ in max-norm sense:
\begin{equation*}
\lambda_{\min}(\Omega_0^{1/2}\Omega^{-1}\Omega_0^{1/2}) \rightarrow 1, \enspace  \lambda_{\max}(\Omega_0^{1/2}\Omega^{-1}\Omega_0^{1/2}) \rightarrow 1, \enspace \nu/\nu_0\rightarrow 1.
\end{equation*}
So we can choose sufficiently small $\varepsilon^\ast_4 >0$ such that the following inequalities hold for all $d(\Omega, \Omega_0) < \varepsilon^\ast_4$,
\begin{equation*}
\begin{split}
B &\leq \frac{\nu_0 + p}{2} \mathbb E \left\vert \log \left[ \frac{\nu + \mathbf y^T \Omega^{-1}\mathbf y}{\nu_0 + \mathbf y^T \Omega_0^{-1}\mathbf y}\right] \right\vert \\
&\leq \frac{\nu_0 + p}{2}\max\left\{ \mathbb E \left\vert \log\left[ \lambda_{\min}(\Omega_0^{1/2}\Omega^{-1}\Omega_0^{1/2}) \wedge \frac{\nu}{\nu_0} \right] \right\vert , \mathbb E \left\vert \log\left[ \lambda_{\max}(\Omega_0^{1/2}\Omega^{-1}\Omega_0^{1/2}) \vee \frac{\nu}{\nu_0} \right] \right\vert \right\} \\
&< \frac{\varepsilon}{6}.
\end{split}
\end{equation*}
Therefore, letting $\varepsilon^\ast = \min\{ \varepsilon^\ast_1, \varepsilon^\ast_2, \varepsilon^\ast_3, \varepsilon^\ast_4\}$, $\vert\nu_0-\nu \vert < \varepsilon^\ast$ and $d_\infty(\Omega_0, \Omega) < \varepsilon^\ast$ imply the following.
\begin{equation*}
\begin{split}
&\text{KL}\Big((\nu_0,\Omega_0), (\nu,\Omega) \Big) \\
&\leq \left\vert \log\left( \frac{\Gamma[(\nu_0 + p)/2 ]}{\Gamma(\nu_0/2) }\nu_0^{\nu_0/2} \right) - \log\left( {\frac{\Gamma[(\nu + p)/2 ]}{\Gamma(\nu/2)}\nu^{\nu/2}} \right) \right\vert \\
&\enspace \enspace + \left\vert \frac{1}{2}\log \left( \det(\Omega)\right) - \frac{1}{2}\log\left(\det(\Omega_0)\right) \right\vert \\
&\enspace \enspace + \left\vert \frac{\nu + p}{2} \mathbb E\log \left[ \nu + \mathbf y^T \Omega^{-1}\mathbf y\right] - \frac{\nu_0 + p}{2} \mathbb E\log \left[ \nu + \mathbf y^T \Omega^{-1}\mathbf y\right] \right\vert \\
&\enspace \enspace + \left\vert \frac{\nu_0 + p}{2} \mathbb E\log \left[ \nu + \mathbf y^T \Omega^{-1}\mathbf y\right] - \frac{\nu_0 + p}{2} \mathbb E\log \left[ \nu_0 + \mathbf y^T \Omega_0^{-1}\mathbf y\right] \right\vert \\
&< \frac{\varepsilon}{3} + \frac{\varepsilon}{3}+ \frac{\varepsilon}{6}+ \frac{\varepsilon}{6} \\
&= \varepsilon 
\end{split}
\end{equation*}
In other words, for any $\varepsilon >0$, we can choose $\varepsilon^\ast >0$ such that
\begin{equation*}
\text{KL}\Big((\nu_0,\Omega_0), (\nu,\Omega) \Big) < \varepsilon, ~ \text{ for all } \vert\nu_0-\nu \vert < \varepsilon^\ast \text{ and }d_\infty(\Omega_0, \Omega) < \varepsilon^\ast.\label{lasteq}
\end{equation*}
Thus Eq.~\ref{WTS} is proved and we have 
\begin{equation*}
\Big\{ (\nu, \Omega) : \vert\nu_0-\nu \vert < \varepsilon^\ast \text{ and }d_\infty(\Omega_0, \Omega) < \varepsilon^\ast \Big\} \subset \Big\{ (\nu, \Omega) : \text{KL}\Big((\nu_0,\Omega_0), (\nu,\Omega) \Big) < \varepsilon \Big\}.
\end{equation*}
The proof of Theorem 2 is done. \qed

\section*{Acknowledgments}

This work was supported by the National Research Foundation of Korea (NRF) grant funded by the Korea government(MSIT) (No. NRF-2020R1A4A1018207).

\bibliographystyle{spbasic}

\bibliography{references}

\end{document}